\documentclass[english]{article}
\usepackage[T1]{fontenc}
\usepackage[latin9]{luainputenc}
\usepackage{geometry}
\usepackage{amstext}
\usepackage{amssymb}
\usepackage{feyn}
\usepackage{tikz}
\usepackage{graphicx}
\usepackage{array}
\usepackage{booktabs}

\usepackage{amsmath}
\usepackage{amsfonts}

\usepackage[normalem]{ulem}
\usepackage{color}
\usepackage{listings,braket}
\usepackage{caption}
\usepackage{subcaption}
\usepackage{float}
\definecolor{darkgreen}{rgb}{0,0.35,0}
\definecolor{Rood}{rgb}{1, 0, 0}

\makeatletter
\usepackage[english]{babel}
\usepackage{feyn}

\makeatother

\begin{document}

\title{\textbf{Comments on  the Bell-Clauser-Horne-Shimony-Holt inequality}}

{\author{\textbf{S.~P.~Sorella}\thanks{silvio.sorella@gmail.com}\\\\\
\textit{{\small UERJ -- State University of Rio de Janeiro,}}\\
\textit{{\small Physics Institute  -- Department of Theoretical Physics   -- Rua S\~ao Francisco Xavier 524,}}\\
\textit{{\small 20550-013, Maracan\~a, Rio de Janeiro, Brazil}}\\
}

\date{}

\maketitle
\begin{abstract}
We discuss the  relationship between the Bogoliubov transformations, squeezed states,  entanglement and  maximum violation of the Bell-CHSH inequality. In particular, we point out that the construction of the four bounded operators entering the Bell-CHSH inequality can be worked out in a simple and general  way, covering a large variety of models, ranging from Quantum Mechanism to relativistic Quantum Field Theories.  Various examples are employed to illustrate the above mentioned  framework. We start by considering  a pair of entangled spin 1 particles and a squeezed oscillator in Quantum Mechanics, moving then to the relativistic complex quantum scalar field and to the analysis of the vacuum state in Minkowski space-time in terms of the left and right Rindler modes. In the latter case, the Bell-CHSH  inequality turns out to be parametrized by  the Unruh temperature. 
\end{abstract}

\section{Introduction}\label{intro}

The aim of this work  is that of pointing out that the  construction of the four Hermitian operators  $A_i$, $B_i$, $i=1,2$  
\begin{equation} 
A^2_i = B^2_i = 1\;, \qquad [A_i, B_k]=0 \;. \label{op} 
\end{equation}
characterizing the quantum violation of the  Bell-Clauser-Horne-Shimony-Holt inequality \cite{Bell:1964kc,Clauser:1969ny,tsi1}, {\it i.e.}
\begin{equation}
| \langle \psi | {\cal C}_{CHSH} | \psi \rangle | = | \langle 
 \psi | (A_1+A_2)B_1 + (A_1-A_2)B_2| \psi \rangle |   > 2 \;,  \label{BCHSH}
\end{equation}
can be achieved in a rather simple and elegant way, which turns out to have general applicability, covering models ranging from Quantum Mechanics to more sophisticated examples such as: relativistic Quantum Field Theories. \\\\In order to illustrate how the setup works, we proceed first by discussing  two examples from Quantum Mechanics. 
\section{Entangled pair of spin 1 particles}\label{spin1}
Let us start by considering  a pair of entangled spin 1 particles. As entangled state $|\psi_s\rangle$, we take the  singlet state:
\begin{equation} 
|\psi_s \rangle = \left( \frac{ |1\rangle_A |-1 \rangle_B - |0 \rangle_A  |0 \rangle_B +
|-1\rangle_A |1\rangle_B}{\sqrt{3}}   \right) \;. \label{singlet}
\end{equation} 
It is easily checked that expression \eqref{singlet} can be thought as the vacuum state of the spin Hamiltonian \begin{equation} 
H = {\vec S}_A \cdot {\vec S}_B = \frac{1}{2} \left( {\vec S}_A + {\vec S}_B \right)^2 - 2 \;. \label{H}
\end{equation}
with $\left({\vec S}_A, {\vec S}_B \right)$ denoting the spin 1 matrices. For the operators $(A_i,B_i)$ we write
\begin{eqnarray} 
A_i |-1\rangle_A & = & e^{i \alpha_i}  |0\rangle_A \;, \qquad A_i |0\rangle_A = e^{-i \alpha_i}  |-1\rangle_A \;, \qquad A_i |1\rangle_A =  |1\rangle_A \;, \nonumber \\
B_i |1\rangle_B & = & e^{i \beta_i}  |0\rangle_B \;, \qquad B_i |0\rangle_B = e^{-i \beta_i}  |1\rangle_B \;, \qquad B_i |-1\rangle_B =  |-1\rangle_B \;, \label{AB1}
\end{eqnarray}
where $(\alpha_i, \beta_i)$ are arbitrary real coefficients. The Hermitian operators $A_i$, $B_i$, $i=1,2$ fulfill the requirement \eqref{op}. \\\\For the Bell-CHSH correlator we obtain  
\begin{equation}
\langle \psi_s | {\cal C}_{CHSH} | \psi_s \rangle = \frac{2}{3} \left( 1 -  cos(\alpha_1+ \beta_1)  -  cos(\alpha_2+ \beta_1) - cos(\alpha_1+ \beta_2) +  cos(\alpha_2+ \beta_2) \right) \;. \label{sp1c}
\end{equation}
Choosing, for example,  $\alpha_2=\beta_2=0$, $\alpha_1= \frac{\pi}{2}$, $\beta_1= \frac{3\pi}{4}$, one gets 
the violation
\begin{equation}
| \langle \Omega | {\cal C}_{CHSH} | \Omega \rangle | = \frac{ 2(2+\sqrt{2})}{3} \approx 2.27 \;,\label{vspin1}
\end{equation}
which compares  well with the value reported by \cite{spin 1}. \\\\The same reasoning applies to the Bell spin $1/2$ singlet state 
\begin{equation} 
|\psi_s \rangle = \left( \frac{ |+\rangle_A |- \rangle_B - |- \rangle_A  |+ \rangle_B }{\sqrt{2}}   \right) \;. \label{bsinglet}
\end{equation} 
For the operators  $(A_i,B_i)$ we have now 
\begin{eqnarray} 
A_i |+\rangle_A & = & e^{i \alpha_i}  |-\rangle_A \;, \qquad A_i |-\rangle_A = e^{-i \alpha_i}  |+\rangle_A \;, \;, \nonumber \\
B_i |+\rangle_B& = & e^{i \beta_i}  |-\rangle_B \;, \qquad B_i |-\rangle_B = e^{-i \beta_i}  |+\rangle_B \;.\label{AB12}
\end{eqnarray}
Setting $\alpha_1=0$, $\alpha_2=\frac{\pi}{2}$, $\beta_1=\frac{\pi}{4}$, $\beta_2=-\frac{\pi}{4}$, one recovers 
Tsirelson's bound \cite{tsi1}, namely 
\begin{equation} 
| \langle \psi_s | {\cal C}_{CHSH} | \psi_s \rangle | = 2\sqrt{2} \;.
\end{equation}

\section{Two degrees of freedom squeezed oscillator}\label{ho} 
In this second example we consider a two degrees of freedom squeezed oscillator. Let us begin  by introducing two annihilation operators $(a,b)$ satisfying 
\begin{eqnarray} 
 \left[ a, a^{\dagger}\right] & = & 1\;, \qquad [a^{\dagger}, a^{\dagger}] =0 \;, \qquad  [a, a] =0 \;, \nonumber \\
 \left[ b, b^{\dagger}\right]  & = & 1\;, \qquad [b^{\dagger}, b^{\dagger}] =0 \;, \qquad  [b, b] =0 \;, \nonumber \\
 \left[a, b \right] & = &  0\;, \qquad [a, b^{\dagger}] =0 \;. \label{ccrqm}
\end{eqnarray} 
The operators $(a,b)$ annihilate the state $|0\rangle$, which is not the vacuum state of the system, {\it i.e.}
\begin{equation} 
a |0\rangle = b |0\rangle = 0 \;. \label{st}
\end{equation}
Let us introduce  the new operators $(\alpha,\beta)$, obtained through the Bogoliubov transformations
\begin{equation}
\alpha = \frac{(a -\eta b^{\dagger})}{\sqrt{1-\eta^2}} \;, \qquad \beta = \frac{(b -\eta a^{\dagger})}{\sqrt{1-\eta^2}}  \;, \label{Bog}
\end{equation} 
where $\eta$ is  a real parameter,  $0<\eta<1$. \\\\It is easily checked that the  operators $(\alpha,\beta)$ fulfill the same commutation relations of eq.\eqref{ccrqm}, namely
\begin{eqnarray} 
 \left[ \alpha, \alpha^{\dagger}\right] & = & 1\;, \qquad [\alpha^{\dagger}, \alpha^{\dagger}] =0 \;, \qquad  [\alpha, \alpha] =0 \;, \nonumber \\
 \left[ \beta, \beta^{\dagger}\right]  & = & 1\;, \qquad [\beta^{\dagger}, \beta^{\dagger}] =0 \;, \qquad  [\beta, \beta] =0 \;, \nonumber \\
 \left[\alpha, \beta \right] & = &  0\;, \qquad [\alpha, \beta^{\dagger}] =0 \;. \label{ccrb}
\end{eqnarray} 
We consider now the normalized squeezed state 
\begin{equation} 
|\eta \rangle = \sqrt{(1-\eta^2)} \;e^{\eta a^\dagger b^\dagger} |0 \rangle \;, \qquad \langle \eta | \eta \rangle = 1 \;. \label{etast}
\end{equation} 
Let us show that the state $|\eta\rangle$ is annihilated by the Bogoliubov operators $(\alpha, \beta)$: 
\begin{equation} 
\alpha |\eta\rangle = \beta |\eta\rangle = 0 \;. \label{zeta}
\end{equation} 
In fact, we have 
\begin{eqnarray} 
( a - \eta b^\dagger) e^{\eta a^\dagger b^\dagger} |0 \rangle & =  & \left[ a, e^{\eta a^\dagger b^\dagger} \right] |0 \rangle - \eta b^\dagger e^{\eta a^\dagger b^\dagger} |0 \rangle  \nonumber \\
& = & \sum_{n=1}^{\infty} \frac{\eta^n}{n!} (b^{\dagger})^n  \left[ a, (a^\dagger)^n  \right]  |0 \rangle - \eta b^\dagger e^{\eta a^\dagger b^\dagger} |0 \rangle \nonumber \\
& = & \eta b^\dagger \sum_{n=1}^{\infty} \frac{\eta^{n-1}}{(n-1)!} (b^{\dagger})^{(n-1)}  (a^{\dagger})^{(n-1)}|0 \rangle - \eta b^\dagger e^{\eta a^\dagger b^\dagger} |0 \rangle = 0 \;. \label{alphaeta}
\end{eqnarray} 
Similarly for the operator $\beta$. As a consequence, the squeezed state $|\eta\rangle$ turns out to be the vacuum state of the Hamiltonian 
\begin{eqnarray} 
H & = & \alpha^\dagger \alpha + \beta^\dagger \beta = \frac{(1+\eta^2)}{(1-\eta^2)} \left( a^\dagger a + b^\dagger b \right) -  \frac{2\eta}{(1-\eta^2)} \left( a^\dagger b^\dagger + a b \right)  + 2 \eta^2. \nonumber \\
H |\eta\rangle & = & 0 \;. \label{H}
\end{eqnarray} 
This expression is nothing but the Hamiltonian worked out  in \cite{Summ} in order to establish the violation of the Bell-CHSH inequality in relativistic free Quantum Field Theory. It becomes apparent now that the role of the parameter $\eta$, $0 < \eta <1$, is that of a coupling constant.

\subsection{Maximum violation of the Bell-CHSH inequality}\label{oscBell}
We are ready now to discuss the violation of the Bell-CHSH exhibited by this model. To introduce the four  Hermitian operators $A_i$, $B_i$, $i=1,2$  
\begin{equation} 
A^2_i = B^2_i = 1\;, \qquad [A_i, B_k]=0 \;,  \label{oposc} 
\end{equation}
we proceed in exactly the same way as done in the case of the spin 1 example. We write the squeezed state $ |\eta\rangle$ as 
\begin{eqnarray} 
|\eta\rangle & = & \sqrt{(1-\eta^2)} \;\sum_{n=0}^{\infty} \eta^n \frac{( a^\dagger b^\dagger)^n}{n!} |0\rangle \nonumber \\
& = & \sqrt{(1-\eta^2)} \;\sum_{n=0}^{\infty} \left( \eta^{(2n)} | (2n_a) (2n_b) \rangle +  \eta^{(2n+1)} | (2n_a+1) (2n_b+ 1) \rangle      \right)  \;, \label{exp}
\end{eqnarray} 
where $| m_a m_b \rangle$ stands for the normalized state 
\begin{equation} 
| m_a m_b \rangle = \frac{(a^\dagger)^m (b^\dagger)^m}{m!} |0\rangle  \;. \label{mab}
\end{equation} 
For the operators $A_i$, $B_i$, $i=1,2$, we have 
\begin{eqnarray} 
A_i | (2n_a) \;m_b \rangle & = & e^{i \alpha_i} | (2n_a+ 1) \;m_b \rangle \;, \qquad A_i | (2n_a+ 1) \;m_b \rangle = e^{-i \alpha_i} | (2n_a)\; m_b \rangle \;, \nonumber \\[2mm]
B_i | m_a\; (2n_b)  \rangle & = & e^{i \beta_i} | m_a \;(2n_b+ 1)  \rangle \;, \qquad B_i | m_a \;(2n_b+ 1)  \rangle = e^{-i \beta_i} | m_a \; (2n_b)  \rangle \;.  \label{ABsq}
\end{eqnarray} 
The operator $A_i$ acts only on the first entry, while $B_i$ only on  the second one. \\\\A quick calculation gives 
\begin{equation} 
\langle \eta | \; A_k B_i \;| \eta \rangle = \frac{2 \eta}{1+\eta^2} \; \cos(\alpha_k + \beta_i) \;. \label{ABcorr}
\end{equation} 
Therefore, for the Bell-CHSH correlator, one finds 
\begin{eqnarray}
 \langle \eta | {\cal C}_{CHSH} | \eta \rangle  & =&   \langle 
 \eta | (A_1+A_2)B_1 + (A_1-A_2)B_2|  \eta \rangle     \nonumber \\
 & = & \frac{2 \eta}{1+\eta^2} \left( \cos(\alpha_1 + \beta_1) + \cos(\alpha_2 + \beta_1) + \cos(\alpha_1 + \beta_2) - \cos(\alpha_2 + \beta_2)  \right) \;.
 \label{BCHSHsq}
\end{eqnarray}
Setting  \cite{Summ}
\begin{equation} 
\alpha_1 = 0\;, \qquad \beta_1 = -\frac{\pi}{4} \;, \qquad \alpha_2 = \frac{\pi}{2} \;, \qquad \beta_2 = \frac{\pi}{4} \;, \label{alpbet}
\end{equation} 
expression \eqref{BCHSHsq} becomes 
\begin{equation} 
\langle \eta | {\cal C}_{CHSH} | \eta \rangle =  2 \;\frac{2 \sqrt{2} \eta}{1+\eta^2} \;, \label{vsq}
\end{equation} 
implying in a violation of the Bell-CHSH inequality whenever  
\begin{equation} 
 \sqrt{2} -1 < \eta < 1 \;. \label{squeezedv}
\end{equation} 
In particular, the maximum violation is attained for values of $\eta \approx 1$, namely 
\begin{equation} 
\langle \eta | {\cal C}_{CHSH} | \eta \rangle \approx   2 \sqrt{2}  \;. \label{mvsq}
\end{equation} 

\section{The relativistic complex scalar quantum Klein-Gordon field}\label{scf}
Let us move now to Quantum Field Theory, by considering the example of a free complex massive scalar Klein-Gordon field, {\it i.e.} 
\begin{equation} 
{\cal L} =   \left( \partial^\mu \varphi^\dagger \partial_\mu \varphi - m^2 \varphi^\dagger \varphi  \right) \;.  \label{cnq1}
\end{equation} 
Expanding $\varphi$ in terms of annihiliation and creation operators, one gets 
\begin{equation} 
\varphi(t,{\vec x}) = \int \frac{d^3 {\vec k}}{(2 \pi)^3} \frac{1}{2 \omega_k} \left( e^{-ikx} a_k + e^{ikx} b^{\dagger}_k \right) \;, \qquad k^0= \omega_k = \sqrt{{\vec{k}}^2 + m^2}  \;, \label{qf}
\end{equation} 
where 
\begin{equation} 
[a_k, a^{\dagger}_q] = [b_k, b^{\dagger}_q] = (2\pi)^3 2\omega_k \delta^3({\vec{k} - \vec{q}}) \;, \label{ccr}
\end{equation}
are the  non vanishing canonical commutation relations. Expression  \eqref{qf} is a too singular object, being in fact an operator valued distribution \cite{Haag:1992hx}. In order to give a well defined meaning to eq.\eqref{qf}, one introduces the smeared operators 
\begin{equation} 
a_f = \int \frac{d^3 {\vec k}}{(2 \pi)^3} \frac{1}{2 \omega_k}  {\hat f}(\omega_k,{\vec k}) \;a_k \;, \qquad 
b_g = \int \frac{d^3 {\vec k}}{(2 \pi)^3} \frac{1}{2 \omega_k} {\hat g}(\omega_k,{\vec k}) \;b_k \;, \label{sm}
\end{equation}
with 
\begin{equation}
{\hat f}(p) = \int d^4x \; e^{-ipx} f(x)  \;, \qquad {\hat g}(p) = \int d^4x \; e^{-ipx} g(x).  \;, \label{fft}
\end{equation} 
where $(f(x),g(x))$ are test functions belonging to the space of compactly supported smooth functions ${\cal C}_{0}^{\infty}(\mathbb{R}^4)$. The support of $f(x)$, $supp_h$, is the region in which the test function $f(x)$ is non-vanishing. \\\\When rewritten in terms of the operators 
$(a_f,b_g)$, the canonical commutation relations \eqref{ccr} read
\begin{equation} 
\left[ a_h,a^{\dagger}_{h'}\right]  = \left[ b_h, b^{\dagger}_{h'}\right] = \langle h | h' \rangle \;, \label{ccrfg}
\end{equation}
where $ \langle h | h' \rangle$ denotes the Lorentz invariant scalar product between the test functions $h$ and $h'$. {\it i.e.} 
\begin{equation} 
\langle h | h' \rangle = \int \frac{d^3 {\vec k}}{(2 \pi)^3} \frac{1}{2 \omega_k}  {\hat h}(\omega_k,{\vec k}) 
{{\hat h}}^{'*} (\omega_k, {\vec k}) = 
\int \frac{d^4 { k}}{(2 \pi)^4} 2\pi \;\theta(k^0) \delta(k^2-m^2)  {\hat h}(k) {\hat h}^{'*}(k)  \;. \label{scpd}
\end{equation} 
In particular, from eq.\eqref{ccrfg}, it follows that 
\begin{equation} 
\left[ a_f,a^{\dagger}_{f}\right]  = \left[ b_f, b^{\dagger}_{f}\right] = ||f||^2   \;, \label{aabb}
\end{equation} 
where $||f||^2 $ is the norm of the test function $f$, {\it i.e.} 
\begin{equation}
||f||^2 = \int \frac{d^4 { k}}{(2 \pi)^4} 2\pi \;\theta(k^0) \delta(k^2-m^2)  {\hat f}(k) {\hat f}^{*}(k)  \;. \label{norm}
\end{equation} 
Let us remark that, by a simple rescaling,  the test functions can be taken to be normalized to 1, namely 
\begin{equation} 
f(x) \rightarrow  \frac{1}{||f||} f(x)  \;, \qquad ||f||=1 \;. \label{norm1}
\end{equation}
Therefore, expression \eqref{aabb} becomes 
\begin{equation} 
\left[ a_f,a^{\dagger}_{f}\right]  = \left[ b_f, b^{\dagger}_{f}\right] = ||f||^2 = 1  \;. \label{aabb1}
\end{equation} 
The vacuum state $|0\rangle$ of the theory is defined as 
\begin{equation} 
a_f |0\rangle = b_g |0\rangle = 0 \qquad \forall \; (f,g) \in {\cal C}_{0}^{\infty}(\mathbb{R}^4) \;. \label{vc}
\end{equation} 
The scalar complex KG field $\varphi$, eq.\eqref{cnq1}, enables us to introduce a squeezed state very similar to that introduced in the previous section, {\it} 
\begin{equation} 
| \sigma \rangle =  \sqrt{(1-\sigma^2)} \;e^{\sigma a^\dagger_f b^\dagger_g} |0 \rangle \;, \qquad \langle \sigma | \sigma \rangle = 1 \;, \qquad 0 < \sigma < 1 \;. \label{sigma}
\end{equation} 
We write thus 
\begin{eqnarray} 
|\sigma\rangle & = & \sqrt{(1-\sigma^2)} \;\sum_{n=0}^{\infty} \sigma^n \;\frac{( a_f^\dagger b_g^\dagger)^n}{n!} |0\rangle \nonumber \\
& = & \sqrt{(1-\sigma^2)} \;\sum_{n=0}^{\infty} \left( \sigma^{(2n)} | (2n_f) (2n_g) \rangle +  \sigma^{(2n+1)} | (2n_f+1) (2n_g+ 1) \rangle      \right)  \;, \label{expqt}
\end{eqnarray} 
where $| m_f m_g \rangle$ stands for the normalized state 
\begin{equation} 
| m_f m_g \rangle = \frac{(a_f^\dagger)^m (b_g^\dagger)^m}{m!} |0\rangle  \;. \label{mabqt}
\end{equation} 
Proceeding as in the  previous example,  the operators $A_i$, $B_i$, $i=1,2$, are given by  
\begin{eqnarray} 
A_i | (2n_f) \;m_g \rangle & = & e^{i \alpha_i} | (2n_f+ 1) \;m_g \rangle \;, \qquad A_i | (2n_f+ 1) \;m_g \rangle = e^{-i \alpha_i} | (2n_f)\; m_g \rangle \;, \nonumber \\[2mm]
B_i | m_f\; (2n_g)  \rangle & = & e^{i \beta_i} | m_f \;(2n_g+ 1)  \rangle \;, \qquad B_i | m_f \;(2n_g+ 1)  \rangle = e^{-i \beta_i} | m_f \; (2n_g)  \rangle \;.  \label{ABsqqt}
\end{eqnarray} 
Again, the operator $A_i$ acts only on the first entry, while $B_i$ only on  the second one. \\\\It turns out that  
\begin{equation} 
\langle \sigma | \; A_k B_i \;| \sigma \rangle = \frac{2 \sigma}{1+\sigma^2} \; \cos(\alpha_k + \beta_i) \;,\label{ABcorqt}
\end{equation} 
so that,  for the Bell-CHSH correlator, one has precisely the expression obtained in the case of the squeezed oscillator, namely 
\begin{eqnarray}
 \langle \sigma | {\cal C}_{CHSH} | \sigma \rangle  & =&   \langle 
 \sigma | (A_1+A_2)B_1 + (A_1-A_2)B_2|  \sigma \rangle     \nonumber \\
 & = & \frac{2 \sigma}{1+\sigma^2} \left( \cos(\alpha_1 + \beta_1) + \cos(\alpha_2 + \beta_1) + \cos(\alpha_1 + \beta_2) - \cos(\alpha_2 + \beta_2)  \right) \;.
 \label{BCHSHsqqt}
\end{eqnarray}
Therefore, using  
\begin{equation} 
\alpha_1 = 0\;, \qquad \beta_1 = -\frac{\pi}{4} \;, \qquad \alpha_2= \frac{\pi}{2} \;, \qquad \beta_2 = \frac{\pi}{4} \;, \label{alpbet}
\end{equation} 
expression \eqref{BCHSHsqqt} becomes 
\begin{equation} 
\langle \sigma | {\cal C}_{CHSH} | \sigma \rangle =  2 \;\frac{2 \sqrt{2} \sigma}{1+\sigma^2} \;, \label{vsqqt}
\end{equation} 
implying in a violation of the Bell-CHSH inequality for
\begin{equation} 
 \sqrt{2} -1 < \sigma < 1 \;. \label{squeezedvqt}
\end{equation} 
The maximum violation is attained for values of $\sigma \approx 1$:  
\begin{equation} 
\langle \eta | {\cal C}_{CHSH} | \eta \rangle \approx   2 \sqrt{2}  \;. \label{mvsqqt}
\end{equation} 
It is worth reminding here that the violation of the Bell-CHSH in free Quantum Field Theory has been established  several decades ago by using Algebraic Quantum Field Theory techniques, see \cite{Summ} and refs.therein. See also \cite{Peruzzo:2022pwv} for a more recent discussion.

\section{Some remarks on the Minkowski vacuum and the Rindler wedges}\label{qft}
Let us conculde with a few remarks on the origin of the violation of the Bell-CHSH inequality  in relativistic Quantum Field Theory. We consider here the case of a real scalar KG field. It can be argued that the violation of the Bell-CHSH inequality can be understood in a simple  way as a consequence of the entanglement properties of the  vacuum state $|0\rangle$, which can be expressed as a squeezed state in terms of left and right Rindler modes \cite{Crispino:2007eb,Harlow:2014yka}, {\it i.e.} 
\begin{equation} 
| 0 \rangle = \prod_i\left( (1-e^{- \frac{2\pi \omega_i}{a}})^\frac{1}{2} \sum_{n_i=0}^{\infty} \; e^{- \frac{\pi n_i \omega_i}{a}} \;  | n_i\rangle_L |n_i\rangle_R  \right) \;, \label{vqft}
\end{equation} 
where $(| n_i\rangle_L, |n_i\rangle_R)$ are the left and right Rindler modes and $T= \frac{a}{2\pi}$ is the Unruh temperature. The relation \eqref{vqft}  follows from the use of a Bogoliubov transformation applied to the  the quantization of the real scalar KG field in the Rindler wedges \cite{Crispino:2007eb,Harlow:2014yka}. Proceeding as in the previous cases,  for the four operators $(A_k,B_k), k=1,2$, we have 
\begin{eqnarray} 
A_k | 2n_i\rangle_L & = & e^{i \alpha_k} | 2n_i + 1\rangle_L \;, \qquad  A_k | 2n_i + 1\rangle_L = e^{-i \alpha_k} | 2n_i\rangle_L \;, \nonumber \\
B_k | 2n_i\rangle_R & = & e^{i \beta_k} | 2n_i + 1\rangle_R \;, \qquad  B_k | 2n_i + 1\rangle_R = e^{-i \beta_k} | 2n_i\rangle_R \\;. 
\end{eqnarray}
As a consequence, setting again  
\begin{equation} 
\alpha_1 = 0 \;, \qquad \alpha_2 = \frac{\pi}{2} \;, \qquad \beta_1 =  -\frac{\pi}{4} \;, \qquad \beta_2=\frac{\pi}{4} \;, \label{vab}
\end{equation} 
the Bell-CHSH inequality can be parametrized as 
\begin{equation} 
| \langle \Omega | {\cal C}_{CHSH} | \Omega \rangle | = 2 {\sqrt{2}} \;\tau(T) \;, \label{tau}
\end{equation} 
where the form factor $\tau  $ reads
\begin{equation} 
\tau(T) = 2 \sum_i \frac{   \left( e^{\frac{\pi \omega_i}{a}} - e^{-\frac{\pi \omega_i}{a} }  \right)    }{ \left( e^{\frac{2\pi \omega_i}{a}} - e^{-\frac{2\pi \omega_i}{a} }  \right)   }  = \sum_i \frac{1}{\cosh(\frac{\omega_i}{2T})}\;, \label{ff}
\end{equation} 
from which it follows that the  violation of the Bell-CHSH inequality in relativistic Quantum Field Theory can be anlyzed in terms of the  Unruh temperature  \cite{prep}.

\section{Conclusion}\label{conclus}

In this work we have pointed out that the four operators $(A_i, B_i), i=1,2$ entering the Bell-CHSH inequality can be constructed in a simple and elegant way. The setup turns out to be of general applicability, ranging from Quantum Mechanics examples to relativistic Quantum Field Theory.  \\\\We have argued that the violation of the Bell-CHSH inequality in relativistic Quantum Field Theory can be understood as a direct consequence of the decomposition of the vacuum sate $|0\rangle$ as a deeply entangled squeezed state in terms of left and right Rindler modes. \\\\These considerations might result in applications in several  Quantum Field Theory models, including Abelian and non-Abelian gauge theories \cite{prep}.

\section*{Acknowledgements}
The authors would like to thank the Brazilian agencies CNPq and FAPERJ for financial support.  S.P.~Sorella is a level $1$ CNPq researcher under the contract 301030/2019-7.

\end{document}